\documentclass[amsfonts,10pt,twocolumn]{revtex4}

\usepackage{epsfig}
\usepackage{dcolumn}

\newcommand{\vb}[1]{{\mathbf{#1}}}
\newcommand{\lb}[1]{\label{#1}}

\newcommand{\bc}{\begin{center}}
\newcommand{\ec}{\end{center}}
\newcommand{\be}{\begin{equation}}
\newcommand{\ee}{\end{equation}}
\newcommand{\bea}{\begin{eqnarray}}
\newcommand{\eea}{\end{eqnarray}}
\newcommand{\ba}[1]{\begin{array}{#1}}
\newcommand{\ea}{\end{array}}

\newcommand{\bt}[1]{\begin{table}[ht]\centering\begin{tabular}{#1}}
\newcommand{\et}[1]{\end{tabular}\caption{\small#1}\end{table}}
\newcommand{\fig}[3]{\begin{figure}[htb]\epsfxsize=80mm\bigskip\centerline{\epsfbox{#1}}\caption{\small\it #2 \label{#3}}\bigskip\end{figure}}

\begin{document}

\thispagestyle{empty}


\vspace{1.5 truecm}

\title{Higher Harmonics in Non-Linear Vacuum from QED Effects\\
Without Low Mass Intermediate Particles}

\author{J. Tito Mendon\c{c}a$^{1)}$}
\author{J. Dias de Deus$^{2)}$}
\author{P. Castelo Ferreira$^{2)}$}
\affiliation{1) CFP and CFIF, Instituto Superior T\'ecnico, 1096 Lisboa, Portugal}
\affiliation{2) CENTRA, Instituto Superior T\'ecnico, Av. Rovisco Pais, 1049-001 Lisboa, Portugal}

\date{\today}

\begin{abstract}
We show that in the presence of a slowly rotating strong transverse magnetic
field there is an infinite spectrum of harmonic wave functions $A_n$ due to
the first order QED correction (in $\alpha^2$) given by the Euler-Heisenberg
Lagrangian. The frequency shifts are integer multiples $\pm \omega_0\,n$ of
the magnetic field angular frequency rotation $\omega_0=2\pi\nu_m$ and the
several modes $n$ are coupled to the nearest harmonics $n\pm 1$. This is a
new effect due to QED vacuum fluctuations, not exploit before, that can
explain, both qualitatively and quantitatively, the recent experimental
results of the PVLAS collaboration without the need of a low mass
intermediate particle, hence may dismiss the recent claim of the discovery
of the axion.
\end{abstract}


\maketitle

In quantum field theory the vacuum is not empty, in the sense that virtual
particle pair creation and annihilation are taking place. Considering the case
of QED -Quantum Electrodynamics- in the presence of external electromagnetic
fields the vacuum shows properties close to the properties of any other
optical medium, as dichroism and birefringence.
These effects can be considered as first order corrections to the Maxwell
theory~\cite{HE,Schwinger,Birula,Adler}, and the basic diagrams, correcting the
photon propagator, are shown in figure~\ref{loops}.
In the presence of low mass neutral particles coupling to two photons, like the
$\pi^0$ and, more interesting, the axion -Primakov effect- or the graviton,
there are additional non-linear effects (see figure~\ref{loops}), which may
be useful in directly detect the axion~\cite{Maiani,massless}.

In this letter we present and exploit an effect not
considered before, in vacuum a strong rotating magnetic
field generates an infinite set of higher harmonics. We  show that
recent experimental results~\cite{PVLA} can be explained in the basis of
pure QED which dismisses the recent claim of the discovery of the
axion~\cite{nature}.

\fig{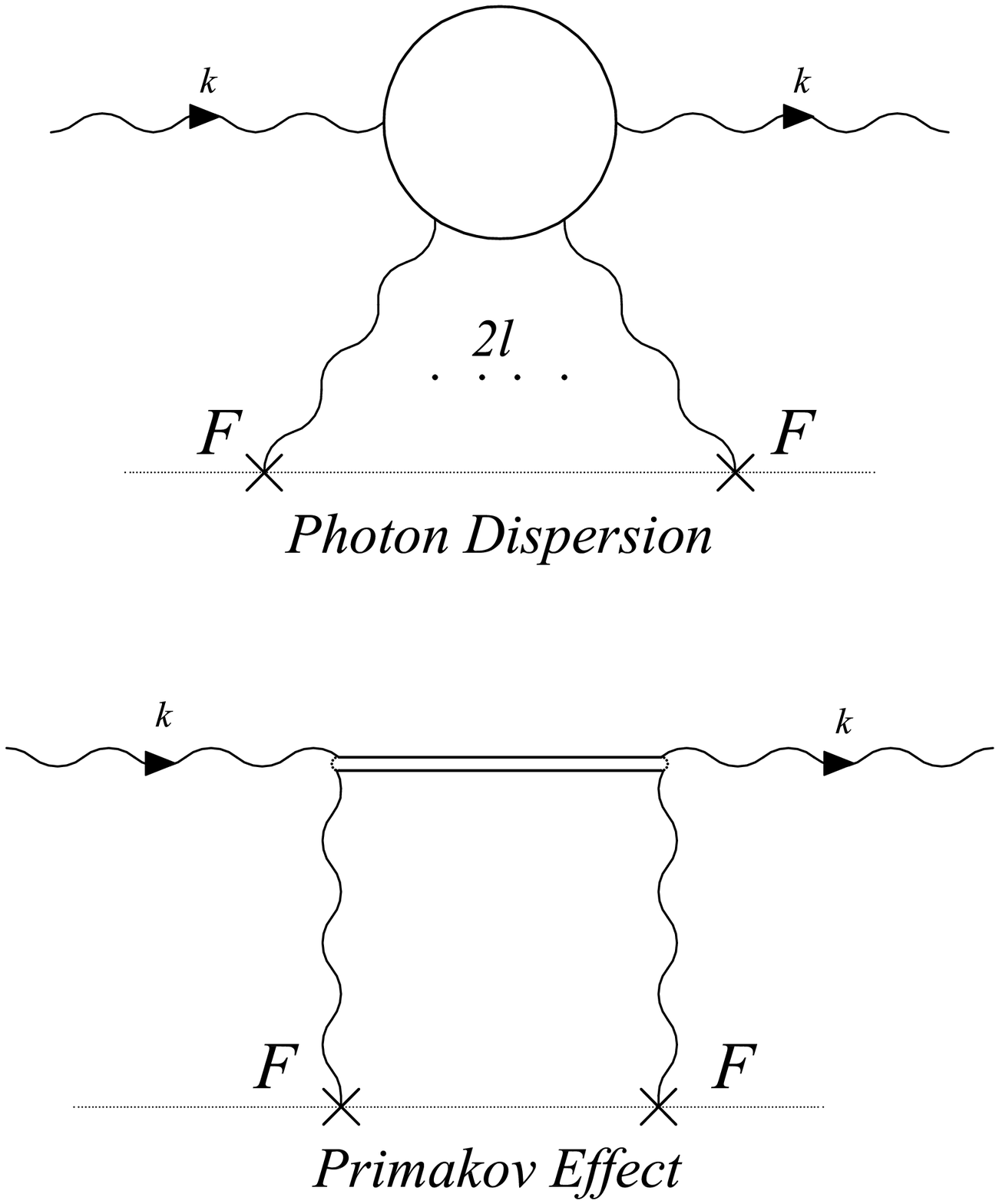}{Feynman diagrams corresponding to the electron loop that contributes to the photon dispersion effect
and the axion exchange.}{loops}

In the QED lowest order correction (radiative corrections in $\alpha^2$) to
the Maxwell Lagrangian ${\mathcal{L}}_{\mathrm{Maxwell}}=-F_{\mu\nu}F^{\mu\nu}$
is given by the regularized Euler-Heisenberg Lagrangian ${\mathcal{L}}^{(2)}$~\cite{HE,Schwinger}
\be
{\mathcal{L}}^{(2)}=\xi\left[4\left(F_{\mu\nu}F^{\mu\nu}\right)^2+7\left(\epsilon^{\mu\nu\delta\rho}F_{\mu\nu}F_{\delta\rho}\right)^2\right]
\ee
where $\xi=2\alpha/(45\,B_c^2)\sim \alpha^2$ and the critical field is
$B_c=m_e^2 c^2/(e\hbar)=4.414\times10^{9}\ T$~\cite{Adler}.
The usual procedure is to take a decomposition of the gauge connection $F$
into an external component $F_0$ plus an internal component
$f_{\mu\nu}=\partial_\mu A_\nu-\partial_\nu A_\mu$, such that we obtain
$F_{\mu\nu}=F_{0,\mu\nu}+f_{\mu\nu}$, and consider the lowest order diagrams
of photon interactions due to the background electromagnetic fields.
There are two main processes that may be considered, \textit{photon splitting}
and \textit{photon dispersion}. Depending on the practical application
and experimental setups there are other processes, such as,
\textit{photon-photon interactions}~\cite{lund} and other photon alternative
processes~\cite{mark} that are relevant. In this work we address only the
process of \textit{photon dispersion} in the low frequency and weak field
limit, i.e. $\omega\ll 2m_e c^2/\hbar$ and $|F|\ll m_e c^2/(\hbar e)$.
The relevant diagrams are the ones containing one internal fermion loop
with two external photons and an even number of exchanged photons between
the fermion loop and the external electromagnetic field
(see figure~\ref{loops}).

We obtain that the relevant radiative corrections to the usual
classical wave equation in order $\alpha^2$ is linear in the photon
field $A$~\cite{Adler}
\be
(\nabla^2-\frac{1}{c^2}\partial_t^2)\vb{A}= \Lambda.\vb{A}+O(\alpha^4)\ ,
\lb{eq}
\ee
where the matrix $\Lambda$ has the eigenvalues~\cite{Birula,Adler}
\be
\lambda_\parallel=14\xi(\omega/c)^2|\vb{Q}|^2\ \mathrm{and}\ 
\lambda_\perp=8\xi(\omega/c)^2|\vb{Q}|^2,
\lb{lambdas}
\ee
where $\vb{Q}=\vb{k}\times\vb{E_0}/|\vb{k}|+\vb{k}\times(\vb{k}\times\vb{B_0})/|\vb{k}|^2$ and
$\vb{k}$ stands for the light wave vector. $\vb{E_0}$ and $\vb{B_0}$ are the
external fields and the directions $\parallel$ and $\perp$ correspond
respectively to the parallel and transverse directions to the external
magnetic field.

We consider linear polarized light traveling in vacuum in the $z$ direction
with wave number $\vb{k}$ and frequency $\omega$ under a slowly rotating
transverse magnetic field rotating with angular frequency
$\omega_0=2\pi\nu_m\ll\omega$ such that we have
$\vb{A}=(A_x,A_y,0)$, $\vb{k}=(0,0,k)$, $w=c\,k$,
$\vb{B}_0(t)=B_0\left[\sin(\omega_0\,t),\cos(\omega_0\,t),0\right]$,
$\vb{E}_0=0$ and $|\vb{Q}|=cB_0$.
In these conditions the gauge field depends only on $z$ and we can write the
equation~(\ref{eq}) for $\vb{A}$ in terms of the parallel direction~($\parallel$)
and transverse direction~($\perp$) as derived by Adler~\cite{Adler}.
However these directions are rotating with an angular frequency $\omega_0$,
therefore we take the directions of the decomposition of the fields to be the
directions of the magnetic field at some initial time $t=0$. Then
the eigenvalues of the matrix $\Lambda$ become time dependent since they are
valid with respect to the magnetic field direction such that we have
\be
\ba{rcl}
\Lambda_\parallel(t)&=&\lambda_\parallel \cos(\omega_0\,t)-\lambda_\perp \sin(\omega_0\,t)\\[5mm]
\Lambda_\perp(t)&=&\lambda_\parallel \sin(\omega_0\,t)+\lambda_\perp \cos(\omega_0\,t)
\ea
\ee
with $\lambda_\parallel$ and $\lambda_\perp$ given in~(\ref{lambdas}).
Considering the above eigenvalues we obtain the field equation
\begin{widetext}
\be
(\partial_z^2-\frac{1}{c^2}\partial_t^2)(A^0_\parallel,A^0_\perp)=\left(\lambda_\parallel \cos(\omega_0\,t)-\lambda_\perp \sin(\omega_0\,t)\right)(A^0_\parallel,0)+\left(\lambda_\parallel \sin(\omega_0\,t)+\lambda_\perp \cos(\omega_0\,t)\right)(0,A^0_\perp)\ ,
\lb{eq_rot}
\ee
\end{widetext}
where the $A^0_\parallel$ and $A^0_\perp$ correspond to the decomposition of
the field along the directions parallel ($\parallel$) and transverse ($\perp$)
to the magnetic field at the initial time $t=0$. For constant magnetic field
($\omega_0=0$) these equations imply that the vacuum acquires a birefringence
due to the existence of two different refractive indices
${\mathtt{n}}_i=1-\lambda_i/2$ along both the directions $i=\parallel,\perp$.
For slowly rotating magnetic fields the usual procedure is to take the zeroth
order expansion of the cosines and sines that hold the same
effect~\cite{massless}. However we show next that even for small $\omega_0$
equation~(\ref{eq_rot}) implies, not just a vacuum birefringence, but instead
a coupling between the neighbor harmonics separated
by frequency shifts of $\pm\omega_0$.
It is important to note that equation~(\ref{eq_rot}) is valid
for slow rotating external fields (see~\cite{Adler} for further details),
i.e. $\omega_0\ll\omega$.

We can cast equation~(\ref{eq_rot}) in complex form by using the definitions
$\lambda_\pm=\lambda_\parallel\pm i\lambda_\perp$, $A_\pm=A_\parallel\pm iA_\perp$
and the usual relations for the trigonometric functions and consider a mode
decomposition of the form
\be
\ba{c}
A_\pm(z,t)=\sum_{n=-\infty}^{+\infty}A^\pm_n(t)\,e^{i\left[k\,z-\omega_n\,t+\theta_n^\pm\right]}\ ,\\[7mm]
\displaystyle c\,k_n=\omega_n=\omega+\omega_0\,n\ ,\ \ \ \theta^\pm_n=\left(-\theta_\lambda\pm\frac{\pi}{4}\right)\,n\ ,\\[5mm]
\displaystyle\lambda_0=\sqrt{\lambda_\parallel^2+\lambda_\perp^2}\ ,\ \ \ \theta_\lambda=\arctan\left(\frac{\lambda_\perp}{\lambda_\parallel}\right)\ .
\ea
\lb{mode_decomp}
\ee

Given the above mode decomposition the differential equation becomes
a recursion relation in the several frequency modes $A_n$
\be
i\,\frac{k}{\sqrt{2}\,c\,\lambda_0}\dot{A}_n=A_{n+1}+A_{n-1}
\lb{eq_rec}
\ee
where $A_n=A_n^+=e^{-i\pi/2}A_n^-$.We neglected a second derivative
term $\ddot{A}_n$ and considered a dispersion relation that renders a
tower of refractive indices in each direction. This is the same
approach of~\cite{Adler} in consistent field approach. Here we have for each
mode $n$ and each direction $i=\parallel,\perp$
a different refractive index
\be
k\,c={\mathtt{n}}^n_i(\omega)\, \omega\ \ ,\ \ {\mathtt{n}}^n_i(\omega)= 1+\frac{\omega_0}{\omega}\,n\ .
\lb{nni}
\ee
The solutions up to an overall phase for the recursive
relation~(\ref{eq_rec}) are modified Bessel function of the first kind
\be
A_n(t)=(-1)^n I_n\left(i\, \tau\right)\ ,\ \ \ \tau=\frac{\sqrt{2}\,c\,\lambda_0\,t}{k}\ .
\lb{sol_A}
\ee
We note that indeed the second derivative of $A_n$ is of order $\alpha^4$
and can be safely neglected. These functions are normalized for any $\tau$
and for even $n$ the functions are pure reals while for odd $n$
the functions are pure imaginary. The only mode that is non
null at $\tau=0$ is $I_0(i\tau)$ such that only for larger values of $t$
the amplitudes of the other modes increase. These properties are important
when imposing the initial conditions, i.e. matching our solution with the
incident wave at the initial time $t=0$.

The results obtain so far are very interesting and describe both vacuum
birefringence and a generalized dichroism in the presence of a strong
rotating magnetic field. The tower of refractive indices~(\ref{nni})
correspond to a birefringence effect quantized in terms of the quantity
$\omega_0/\omega$. The Bessel functions solutions account for a generalized
dichroism effect that decreases the amplitude of the incident wave in one
mode and increases it in the nearest modes, we note that the QED effects
considered here do preserve momenta such that there is no real loss of
energy, it simply is transfered from one mode to the nearest modes.
These effects can be understand as follows, from equation~(\ref{eq_rec})
we see that the variation of each mode amplitude $A_n$ over time contributes
to the nearest modes $A_{n\pm 1}$, this accounts for the generalized dichroism
effect in the sense that no real absorption happens, therefore we have a
\textit{true} rotation of polarization (in the sense considered
in~\cite{PVLA}). As for birefringence we note that by combining the
nearest modes (e.g. $n_\parallel=n$ and $n_\perp=n\pm 1$) in different
directions we will have a net birefringence effect given by
${\mathtt{n}}_\perp-{\mathtt{n}}_\parallel=\pm \omega_0/\omega$ such
that each mode combination correspond to a wave with
elliptic polarization. It is also interesting to note that,
although the description for the propagating wave is given in terms
of the collective field $A$, we may as well interpret these
solutions as cross section for photons. In such interpretation
the allowed configurations for a single photon are independent
in both spatial dimensions ($\parallel$,$\perp$), such that
indeed we have elliptic polarized waves.

A solution for the electric field compatible with linearized polarized waves
at the initial time $\tau=0$ is
\begin{widetext}
\be
\vb{E}=E_0 \left\{\sum_{n=-\infty}^{+\infty}\left[\,I_{2n}\left(i\,\tau\right)\cos\left(kz-\omega_{2n}t+\theta_{2n}\right)-i\,I_{2n-1}\left(i\tau\right)\sin\left(kz-\omega_{2n-1}\,t-\theta_{2n-1}\right)\,\right]\right\}\left(\sin(\theta_0),\cos(\theta_0)\right)\ .
\lb{solution}
\ee
\end{widetext}
Here $E_0$ is the electric field amplitude, $\omega_n=\omega+\omega_0\,n$
and $\theta_n=-(\theta_\lambda+\pi/4)\,n$. This field is defined
in the referential of the parallel and transverse direction to the
magnetic field at the initial time $t=0$ such that $\theta_0$ stands
for the angle between the original linear polarization direction and the
magnetic field direction at $t=0$.

Take the $n_\parallel=n_1$ mode in one of the directions together with one of
the other modes $n_\perp=n_2>n_1$ in the other direction such that this
combination has now an elliptic polarization and the difference of argument
of the trigonometric functions corresponding to each direction correspond to a
phase difference of
$\Delta\varphi_{n_2-n_1}=(n_2-n_1)(\omega_0\,t+\theta_\lambda+\pi/4)$.

We proceed to compare our results with the recent experimental results of the
PVLAS collaboration~\cite{PVLA}. For the apparatus of that experiment an
incident polarized laser beam traverses a region under a strong rotating
magnetic field such that it is possible to measure polarization rotations
(that correspond to vacuum dichroism) and ellipticities (that correspond to
vacuum birefringence).

We can write the amplitude measured in the experience for
our solutions as
\be
\ba{rcl}
\displaystyle{\mathcal{I}}&=&\displaystyle\left[2\sum_{n=1}^{+\infty}\Gamma_n\,|I_n(i\,\tau)|\,\cos(n\,(\omega_0\,t+\theta_\lambda+\theta_0+\pi/4))\right.\\[5mm]
             & &\displaystyle\left.\ \ \ +\eta_0\cos(\omega_S\,t+\theta_S)+\Gamma_0\vphantom{\sum_{n=1}^{+\infty}}\right]^2\ .
\ea
\lb{I}
\ee
Here $\eta_0\,\cos(\omega_S\,t+\theta_S)$ stands for the modulator
wave~\cite{PVLA} and heterodyne detection is used to identify the several
harmonics present due to rotations. $\Gamma_n$ are corrections that encode
uncompensated dispersions for the several modes $n$ due, either to systematic
deviations, or unaccounted processes. In~(\ref{I}) we sum over rotations in
one direction only, that is why we have a factor of 2 in front of the sum. By
expanding the square of this expression we obtain the angular frequency
spectrum and the respective amplitudes and phases for each of them. We list
the more significant frequencies that can be measured near the modulator
frequency $\omega_S$ and list them in table~\ref{freq}.
\begin{table}
\bea
\ba{c|cc}
\mathrm{Frequency}     &\mathcal{I}          &\mathrm{Phase}\\ \hline
0                      &\displaystyle \Gamma_0^2+2\sum_{n=1}^{+\infty}(\Gamma_n\,I_n)^2+\frac{\eta_0^2}{2}&--\\[5mm]
\omega_S               &2\Gamma_0\eta_0       &\theta_S\\[5mm]
\omega_S\pm\omega_0\,n &2\Gamma_nI_n\eta_0    &\theta_S\pm n\,(\theta_\lambda+\theta_0+\pi/4)
\ea
\nonumber
\eea
\caption{The relevant angular frequency spectrum measured near the modulator frequency
$\omega_S$ with respective amplitudes and phases. There is an infinite set of other
frequencies that have lower amplitudes.
\lb{freq}}
\end{table}

We take the values of figure~2 of the original reference~\cite{PVLA}
(expressed in I.S. units)
\be
\ba{rclcrcl}
\omega&=&1.772\times 10^{+15}\ s^{-1}&\ \ \ \ &\omega_0&=&0.7\pi\ s^{-1}\\[5mm]
\Delta z&=&c\Delta t=4.6\times 10^{+4}\ m&\ \ \ \ &B_0&=&5.5\ T\ .
\ea
\lb{exp_cond}
\ee
The theoretical results obtained above are valid for these experimental
conditions, in particular the magnetic field is slowly rotating
($\omega_0\ll \omega$)and obey the low frequency and weak field
approximations, respectively
$\omega\sim 10^{15}\ll 2m_e\,c^2/\hbar\sim 10^{21}$
and $B_0^2\sim 10\ll m_e\,c^2/(e\hbar) \sim 10^{39}$.
For the values given in~(\ref{exp_cond}) and using
the finesse of the Fabry-Perot cavity ${\mathcal{F}}\approx 7\times 10^4$
(corresponding to an increase of optical path
$N\approx 4.46\times 10^4$)~\cite{PVLA} we have
from~(\ref{sol_A}) with $\tau=0.0088$~\cite{Birula} that
$|I_0(i\,\tau)|\approx 0.9998$,
$|I_1(i\,\tau)|\approx 4.384\times 10^{-3}$,
$|I_2(i\,\tau)|\approx 9.609\times 10^{-6}$ and
$|I_3(i\,\tau)|\approx 1.404\times 10^{-8}$.
We use these values to fit our solutions to the data expressed
in figure~2 of reference~\cite{PVLA}. We considered that the
background noise is $\approx-135\ dBVrms$ and assumed values for
$\eta_0$ and the $\Gamma_i$'s that allow us to fit the experimental
results. These values are given as an example and are not based in
a detailed analysis. In table~\ref{table} we list a possible set
of values (in dBVrms) that fits the experimental
results of~\cite{PVLA}.
\begin{table}
\bea
\ba{c|cccccc}
\mathrm{frequency}&\ \ \ &{\mathcal{I}}_{\mathrm{exp}}&\ \ \ &\eta_0I_n&\ \ \ &\Gamma_i\\
(Hz)&\ \ \ &(dBVrms)&\ \ \ &(dBVrms)&\ \ \ & \\[2mm]\hline
 & & & &\\   
506        & &\approx -51 & &--& &\approx +16.93\\[5mm]
506\pm 0.35& &\approx -91 & &\approx -90.70  & &\approx +0.704\\[5mm]
506\pm 0.70& &\approx -118& &\approx -117.30 & &\approx -0.704\\[5mm]
506\pm 1.05& &< -135      & &\approx -145.65 & &--
\ea
\nonumber
\eea
\caption{We list the experimental values taken from figure~2 of
reference~\cite{PVLA} with the theoretical values given by the
$I_n(i\tau)$ up to $n=3$. We considered $\eta_0\approx -68.63\ dBVrms$.
This value, as well as the value of the $\Gamma_n$'s, are only given as
an example that it is possible to fit the data to our solutions.
The theoretical value found for the amplitude of a rotation corresponding
to $n=3$ is $\approx-145.65\ dBVrms$ and is therefore below the noise level
of the experiment that we considered ($\approx-135\ dBVrms$).\lb{table}}
\end{table}
We note that the value obtained for the mode $n=3$ ($\approx -145.65\ dBVrms$)
is below the noise level ($\approx-135\ dBVrms$), hence it could not
be detected for the available data.

To conclude we compute the expected rotation values corresponding
to the second harmonic $n=2$.
By direct geometrical analysis we conclude that the angular rotations
of polarization are given by the generic expression
$\varphi=\theta_0-\arctan\left(\frac{E_\perp^{\mathrm{out}}}{E_\parallel^{\mathrm{out}}}\right)$,
where $\vb{E}^{\mathrm{out}}=(E_\parallel^{\mathrm{out}},E_\perp^{\mathrm{out}})$
stands for the radiation electric field reaching the detector.
For a given harmonic $n$ we have the possible combinations
$\vb{E}^{\mathrm{out}+}_n=[I_0\sin(\theta_0)\cos(\delta(t,z)),I_n\cos(\theta_0)\cos(\delta(t,z)+\omega_0\,\Delta t+\arctan(7/4)+\pi/4)]$ and
$\vb{E}^{\mathrm{out}-}_n=[I_n\cos(\theta_0)\cos(\delta(t,z)+\omega_0\,\Delta t+\arctan(7/4)+\pi/4),I_0\sin(\theta_0)\cos(\delta(t,z))]$.
Here the phase $\delta(t,z)=k\,z-\omega\,t$ is the phase of the wave
when reaching the detector and the sine and cosine of
$\theta_0$ is due to the angle between the original polarization
and the magnetic field when the radiation enters the magnetic field
as expressed in the solution~(\ref{solution}).

Based in the values~(\ref{exp_cond}) with $\theta_0=\pi/4$ (that corresponds to
the maximum effect) and assuming $\delta(t,z)=0$ we obtain for the rotations
corresponding to n=2 $\varphi_{n=2}^+\approx 3.78\times 10^{-6}\,rad$. This value
is one order of magnitude higher than the experimental values measured
($\sim 10^{-7}$). A possible explanation for this result may be due to
a different initial value of the angle between the polarization and the
magnetic field, $\theta_0$. If we consider for instance $\theta_0=\pi/48$
we obtain $\varphi_{n=2}^+\approx 2.48\times 10^{-7}\,rad$ which
is closer to the experimental observed rotation.
Other possibilities may be due to an effective lower path length
due either to decoherence or unaccounted losses.

So we have presented a new effect of vacuum polarization rotation due
to external strong magnetic fields based in pure QED.
This effect generates an infinite number of harmonics
and explains, both qualitatively and quantitatively, the
PVLAS experimental results~\cite{PVLA}. These experimental results show
that the PVLAS experiment gives (as far as we know for the first time)
a beautiful demonstration of the QED properties of vacuum and have before
hand registered an effect that have not been justified at theoretical level.
Our results may dismiss the recent claim of the finding of the axion~\cite{nature}.
Both for future use in experiments currently
being planned and for a correct interpretation of already
detected (and to be detected) physical phenomena~\cite{11}, a
full analysis of existent data is required. In particular, we
note that, if the noise level can be decreased, the third peak
($n=\pm 3$) should be detected.

{\bf Erratum}

A new mechanism was proposed to explain the observed spectrum in the PVLAS experiment~\cite{PVLA}.
We have predicted a  series of new signals with frequencies $\omega + n \omega_0$, for $n$ integer, where $\omega$ is the laser frequency and $\omega_0$ the angular frequency of the rotating magnetic field. But two mistakes were made in this work. First, the secondary signals have frequencies $\omega + 2 n \omega_0$. Second, our numerical estimates are wrong by a factor of $10^3$~\cite{Adler_2}, and therefore cannot explain the observed signal. It should be stressed that signals at frequencies $\omega + n \omega_0$ can still be generated, but they are not relevant to the PVLAS experiment.

\ \\
{\large\bf Acknowledgements}

Work of PCF supported by SFRH/BPD/17683/2004.


\begin{thebibliography}{99}

\bibitem{HE} W. Heisenberg and H. Euler, Z. Physik {\bf 98} (1936) 714, \texttt{physics/0605038}.
\bibitem{Schwinger} J. Schwinger, Phys. Rev. {\bf 82} (1951) 664.

\bibitem{Birula} Z. Bialynicka-Birula and I. Bialynicka-Birula, Phys. Rev. {\bf D2} (1970) 2341.

\bibitem{Adler} S. L. Adler, Ann. Phys. {\bf 67} (1971) 559.

\bibitem{Maiani} L. Maiani, R. Petronzio and E. Zavattini, Phys. Lett. {\bf B175} (1986) 359.

\bibitem{massless} G. Raffelt and L. Stodolski, Phys. Rev. {\bf D37} (1988) 1237.

\bibitem{PVLA} E. Zavattini and al., Phys. Rev. Lett. {\bf 96} (2006) 110406.

\bibitem{lund} E. Lundstr\"om et al., Phys. Rev. Lett. {\bf 96} (2006) 083602.

\bibitem{mark} M. Marklund, P. K. Shukla, Rev. Mod. Phys. {\bf 78} (2006) 591.

\bibitem{nature} S. Lamoreaux, Nature Vol. {\bf 441} (2006) 31.

\bibitem{11} E. Mass\'o, \texttt{hep-ph/0607215}; Y. N. Gnedin, M.Y. Piotrovich, and T. M. Natsvlishvili, \texttt{astro-ph/0607294};
U. Kotz, A. Ringwald, and T. Tschentscher, \texttt{hep-ex/0606058}; E. Mass\'o and J. Redondo, \texttt{hep-ph/0606163};
A.V. Afanasev, O. K. Baker, and K.W. McFarlane, \texttt{hep-ph/0605250}.

\bibitem{Adler_2} S. L. Adler, hep-ph/0611267.

\end{thebibliography}
\end{document}